\documentclass{webofc}
\input{latex/packages.tex}
\newcommand{\HiFa}{\texttt{HistFactory}}

\newcommand{\pyhf}{\texttt{pyhf}}

\newcommand{\funcX}{\texttt{funcX}}

\newcommand*{\TeV}{\ensuremath{\text{Te\kern -0.1em V}}}
\newcommand*{\GeV}{\ensuremath{\text{Ge\kern -0.1em V}}}
\newcommand*{\MeV}{\ensuremath{\text{Me\kern -0.1em V}}}
\newcommand*{\keV}{\ensuremath{\text{ke\kern -0.1em V}}}
\newcommand*{\eV}{\ensuremath{\text{e\kern -0.1em V}}}

\newcommand*{\ifb}{\mbox{fb\(^{-1}\)}}

\begin{document}
\title{Distributed statistical inference with \pyhf{} enabled through \funcX{}}

\author{\firstname{Matthew} \lastname{Feickert}\inst{1}\fnsep\thanks{\email{matthew.feickert@cern.ch}} \and
 \firstname{Lukas} \lastname{Heinrich}\inst{2}\fnsep\thanks{\email{lukas.heinrich@cern.ch}} \and
 \firstname{Giordon} \lastname{Stark}\inst{3}\fnsep\thanks{\email{giordon.holtsberg.stark@cern.ch}} \and
 \firstname{Ben} \lastname{Galewsky}\inst{4}\fnsep\thanks{\email{bengal1@illinois.edu}}%
 \setcounter{footnote}{-1}% Don't show footnote asterisk
 \footnote{\\Reproduction of this article or parts of it is allowed as specified in the CC-BY-4.0 license.}
}

\institute{University of Illinois at Urbana-Champaign, Urbana, IL, USA
 \and
 CERN, Geneva, Switzerland
 \and
 University of California Santa Cruz SCIPP, Santa Cruz, CA, USA
 \and
 National Center for Supercomputing Applications, Urbana, IL, USA
}

\abstract{%
 In High Energy Physics facilities that provide High Performance Computing environments provide an opportunity to efficiently perform the statistical inference required for analysis of data from the Large Hadron Collider, but can pose problems with orchestration and efficient scheduling.
The compute architectures at these facilities do not easily support the Python compute model, and the configuration scheduling of batch jobs for physics often requires expertise in multiple job scheduling services.
The combination of the pure-Python libraries \pyhf{} and \funcX{} reduces the common problem in HEP analyses of performing statistical inference with binned models, that would traditionally take multiple hours and bespoke scheduling, to an on-demand (fitting) ``function as a service'' that can scalably execute across workers in just a few minutes, offering reduced time to insight and inference.
We demonstrate execution of a scalable workflow using \funcX{} to simultaneously fit 125 signal hypotheses from a published ATLAS search for new physics using \pyhf{} with a wall time of under 3 minutes.
We additionally show performance comparisons for other physics analyses with openly published probability models and argue for a blueprint of fitting as a service systems at HPC centers.

}
\maketitle
\section{Introduction}\label{sec:introduction}

Researchers in High Energy Physics (HEP) and other fields are encouraged by their funding bodies to take advantage of the High Performance Computing (HPC) facilities constructed at various institutions.
These facilities include capable machines such as Theta at Argonne National Laboratory with 280,000 cores and 192 hardware-accelerated GPUs~\cite{ThetaANL}.
While powerful, these architectures do not easily support the Python compute model.
Users must construct batch jobs and submit them to a queue for execution when compute time is available.
The results are stored on the file system and must be stitched back together once all of the jobs have completed.
On many of these systems, Python tooling lags the current state of the art and configuring modern Python libraries to use HPCs can be a tedious task and require expertise.\\

In HEP a core component of analysis of data collected at the Large Hadron Collider (LHC) is performing statistical inference for binned models to extract physics information.
The statistical fitting tools used in HEP have traditionally been implemented in C++, but in recent years \pyhf{}~\cite{pyhf,pyhf_joss}, a pure-Python library with automatic differentiation and hardware acceleration, has grown in use for analysis related statistical inference problems.
The fitting of multiple different hypotheses for new physics signatures (signals) is a computational problem that lends itself easily to parallelization, but is hampered on HPC environments by the additional tooling overhead required, which can be very difficult to master.
Through use of \funcX{}~\cite{funcX_paper}, a pure-Python high performance function serving system designed to orchestrate scientific workloads across heterogeneous computing resources, \pyhf{} can be used as a highly scalable (fitting) function as a service (FaaS) on HPCs.

\section{Fitting as a Service Methods and Technologies}\label{sec:methods}
\subsection{\pyhf{}}\label{subsec:pyhf}

For measurements in HEP based on binned data (histograms), the \HiFa{}~\cite{Cranmer:1456844} family of statistical models has been widely used for likelihood construction in Standard Model measurements (e.g. Refs.~\cite{HIGG-2013-02,Aaij:2015sqa}) as well as searches for new physics (e.g. Ref.~\cite{SUSY-2016-10}) and reinterpretation studies (e.g. Ref.~\cite{Alguero:2020grj}).
\pyhf{} is a pure-Python implementation of the \HiFa{} statistical model for multi-bin histogram-based analysis.
\pyhf{}'s interval estimation is computed through either the use of the asymptotic formulas of Ref.~\cite{Cowan:2010js} or empirically through pseudoexperiments (``toys'' in HEP parlance).
Through adoption of open source ``tensor'' computational Python libraries (i.e. NumPy, TensorFlow, PyTorch, and JAX), \pyhf{} is able to leverage tensor calculations to outperform the traditional C++ implementations of \HiFa{} on data from real LHC analyses.
\pyhf{} can additionally leverage automatic differentiation and hardware acceleration from the tensor libraries that support them to further accelerate fitting.
Through use of JSON to provide a declarative plain-text serialisation for describing \HiFa{}-based likelihoods~\cite{ATL-PHYS-PUB-2019-029} --- well suited for reinterpretation and long-term preservation in analysis data repositories such as HEPData~\cite{Maguire:2017ypu} --- \pyhf{} has also become a widely used tool across experiment and theory.
Given its lightweight core dependencies and wide distribution through The Python Package Index (PyPI), Conda-forge, and CernVM File System (CernVM-FS) it is easily installable on a wide variety of platforms, including Linux containers.
Minimally sized Docker images containing stable releases of \pyhf{} are also distributed through Docker Hub.

\subsection{\funcX{}}\label{subsec:funcX}
\funcX{} is a distributed FaaS platform designed to support the unique needs of scientific computing.
It combines a reliable and easy-to-use cloud-hosted interface with the ability to securely execute functions on distributed endpoints deployed on various computing resources.
\funcX{} supports many high performance computing systems and cloud platforms, can use three popular container technologies, and can expose access to heterogeneous and specialized computing resources.
The \funcX{} API is a powerful tool to developers and analysts, allowing servable functions to be created from arbitrary Python functions.
To execute a remote function registered with an instance of the \funcX{} client class, a function on the \funcX{} client is called and passed the remote function's required arguments, as seen in~\Cref{lst:funcX_registration_example}.

\begin{listing}
\begin{minted}{python}
import json
from pathlib import Path
from time import sleep

from funcx.sdk.client import FuncXClient
from pyhf.contrib.utils import download


def prepare_workspace(data):
    import pyhf

    return pyhf.Workspace(data)


if __name__ == "__main__":
    # locally get pyhf pallet for analysis
    if not Path("1Lbb-pallet").exists():
        download("https://doi.org/10.17182/hepdata.90607.v3/r3", "1Lbb-pallet")
    with open("1Lbb-pallet/BkgOnly.json") as bkgonly_json:
        bkgonly_workspace = json.load(bkgonly_json)

    # Use privately assigned endpoint id
    with open("endpoint_id.txt") as endpoint_file:
        pyhf_endpoint = str(endpoint_file.read().rstrip())

    fxc = FuncXClient()

    # Register function and execute on worker node
    prepare_func = fxc.register_function(prepare_workspace)
    prepare_task = fxc.run(
        bkgonly_workspace, endpoint_id=pyhf_endpoint, function_id=prepare_func
    )

    # Wait for worker to finish and retrieve results
    workspace = None
    while not workspace:
        try:
            workspace = fxc.get_result(prepare_task)
        except Exception as excep:
            print(f"prepare: {excep}")
            sleep(10)
\end{minted}
 \caption{Truncated example of use of the \funcX{} Python API to register and execute a \pyhf{} function on a \funcX{} endpoint and then retrieve the execution output.
 This example shows evaluation of the background only hypothesis workspace and is extended in a similar fashion to evaluate the signal hypothesis workspaces.}
 \label{lst:funcX_registration_example}
\end{listing}

A \funcX{} endpoint is a logical entity that represents a compute resource.
The endpoint is managed by an agent process that allows the \funcX{} service to dispatch functions to that resource for execution.
The agent handles authentication and authorization, provisioning of nodes on the compute resource, and monitoring and management.
Administrators or users can deploy a \funcX{} agent and register an endpoint for themselves or others, providing descriptive  metadata (e.g. name, description).
As seen in~\Cref{lst:funcX_registration_example}, each endpoint is assigned a unique identifier for subsequent use.\\

Behind the scenes, \funcX{} uses a heterogeneous executor model based on the Parsl parallel scripting project~\cite{Parsl_paper}.
This architecture uses manager processes which run at a particular compute site.
The managers are configured to use one of many different task execution providers, such as HTCondor, Slurm, Torque, and Kubernetes.
With this architecture it is possible to launch tasks on any of these different environments using the same, simple invocation syntax.
Resources on different HPCs can be accessed by simply changing the endpoint identifier.
The endpoint's configuration has numerous settings to tune the endpoint's use of compute resources to the specific environment and the computational profile of the job at hand.
This can include configuring workers to take advantage of small windows of CPU availability, or allowing the workers to wait for a larger allocation to be available.
In either event, the \funcX{} service will cause the task to wait and execute as many tasks as it can when the workers are available.
This helps to match the job profiles against a wide variety of compute environments.
The endpoint process itself is light weight and consumes minimal resources while awaiting new tasks to schedule on workers.\\

The dependencies required to execute user defined functions can be setup in multiple ways.
Developers can provide a command to install dependencies that will be executed on each worker prior to scheduling any tasks (e.g. \texttt{pip install "pyhf[contrib]"}).
Environments that support containerization through Shifter or Singularity can specify a container in the setup.
This is easiest to administer; however, it requires that all tasks running on that endpoint only depend on these provided settings.
Currently, the Kubernetes executor offers more sophisticated support for containers.
Users may register a Docker image with \funcX{} and associate that image with a function.
The Kubernetes executor will launch worker pods with the requested container as needed to support task invocations.

\subsection{Current and Future FaaS Analysis Facilities}\label{subsec:FaaS_analysis_facilities}

Through the capabilities of \funcX{} and the fitting performance and declarative nature of \pyhf{} there is opportunity to create a fitting FaaS analysis facility blueprint for leveraging the scaling potential of HPC centers and dedicated hardware acceleration resources.
The blueprint can then be replicated in deployment at HPC centers with available resources and allocation.
\Cref{fig:infrastructure_perspective} shows possible cyberinfrastructure and system design prospects, from the viewpoints of developers and users, to create a deployment of the blueprint.
Through the development of \pyhf{} and \funcX{} and through this work, the authors have implementations of the ``Development'', ``Building'', and ``Deploying'' stages of the ``FaaS Team'' section of \Cref{fig:infrastructure_perspective} as well as the ``Fit'' stage of the ``End Users''.
The remaining critical infrastructure and administrative stages to create a functional FaaS analysis facility do not have existing implementations at the time of writing (2021), but are the subject of ongoing discussions inside of the Institute for Research and Innovation in Software for High Energy Physics (IRIS-HEP)~\cite{IRIS-HEP:strategic-plan}.
As a demonstration of the ability to reduce the time to insight such facilities would offer, we use the RIVER HPC system's~\cite{RIVER_HPC} deployment of \funcX{} to simultaneously evaluate the 125 signal hypothesis patches from the published analysis of a search for electroweakinos with the ATLAS detector using the full Run-2 dataset of \(139~\ifb\) of \(\sqrt{s} = 13\,\text{TeV}\) proton-proton collision data~\cite{SUSY-2019-08} with \pyhf{}.
RIVER is able to use \funcX{}'s Slurm task execution provider in concert with a Docker image containing all runtime dependencies and the Kubernetes \funcX{} executor to leverage the 120 VM cluster for batch jobs.
Each pair of VMs share a hardware node with two Intel Xeon E2650 v3 processors (24 cores), 16 x 16GB TruDDR4 Memory (256GB), two 800GB SATA MLC SSD's (1.6TB), and a 10GigE network --- providing an excellent testing grounds for scaling workflows.
\clearpage

\begin{figure}[!htpb]
    \centering
    \includegraphics[width=\textwidth]{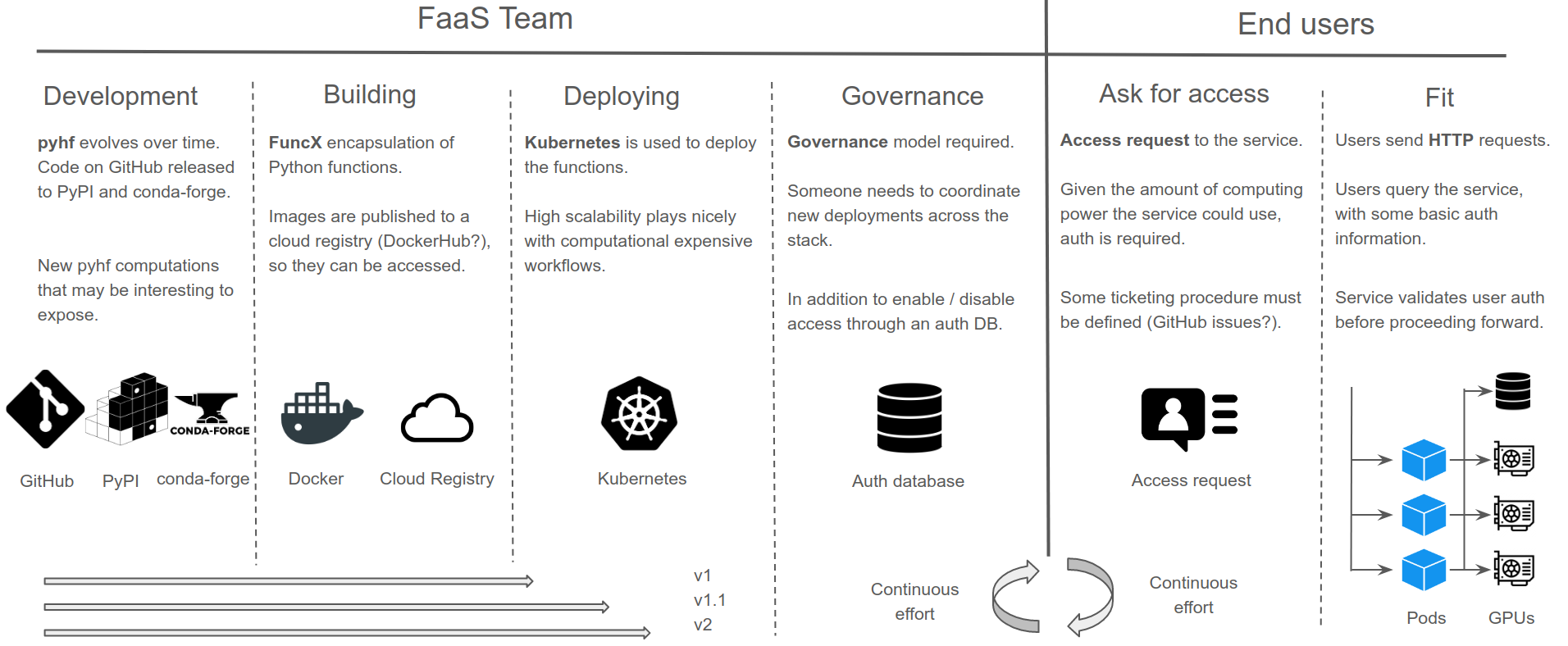}
    \caption{Example infrastructure design from the developer and user perspectives for a \pyhf{} and \funcX{} based fitting FaaS system for physics analysis.~\cite{portable_inference_workshop}}
    \label{fig:infrastructure_perspective}
\end{figure}

\section{Scaling of Statistical Inference}\label{sec:results}
Using the \funcX{} configuration deployed on RIVER described in~\Cref{subsec:FaaS_analysis_facilities}, \funcX{} is able to receive posted JSON serializations of the \pyhf{} pallet containing the background workspace and signal patches downloaded from HEPData~\cite{ATLAS_SUSY_1Lbb_pallet}, start \funcX{} worker nodes, send patched workspaces to each worker, fit the workspace and return the results with a user wall time of under 3 minutes.
As this wall time includes data transfer to and from the user's machine and RIVER, and worker node orchestration time, the time required for inference alone is even smaller.
Example typical run output and performance can be seen in~\Cref{lst:funcX_demo_output}.
The timing results over multiple trials for Ref.~\cite{ATLAS_SUSY_1Lbb_pallet}, using \pyhf{}'s NumPy backend and SciPy optimizer, along with the results from additional analyses~\cite{SUSY-2018-09,SUSY-2018-04} that have openly published probability models as \pyhf{} pallets on HEPData~\cite{ATLAS_SUSY_SS3L_pallet,ATLAS_SUSY_staus_pallet}, are summarized in~\Cref{table:performance} and visualized in~\Cref{fig:timing_barplot_river} and compared to the fit time for all patches on a single node.
All code used in these studies is publicly available on GitHub at Ref.~\cite{study_code,study_code_zenodo_doi}.\\

As \funcX{} endpoints run as users on the resources they are deployed on, and do not have elevated privileges, the number of worker nodes available is not an endpoint configurable option and so is not reported in this work.
Endpoints will utilize available resources effectively and allocate jobs to any available workers given their configuration settings.
A typical way to parameterize the range of available workers that an endpoint can scale work out on is by the \funcX{} endpoint configuration variables \texttt{max\_blocks} and \texttt{nodes\_per\_block} that control the available compute blocks --- the basic unit of resources acquired from an execution provider (e.g. a Slurm scheduler).
\texttt{max\_blocks} controls the maximum number of blocks that can be active per \funcX{} executor and \texttt{nodes\_per\_block} controls the number of nodes requested per block~\cite{Parsl_paper}.
These configuration parameters determine for a given value (generally $1$) of \texttt{parallelism} --- the ratio of task execution capacity to the sum of running tasks and available tasks --- how \funcX{} provisions blocks and distributes work to nodes.
The results summarized in~\Cref{table:performance} use $\texttt{max\_blocks}=4$ and $\texttt{nodes\_per\_block}=1$.
\clearpage

\begin{table}[htpb]
\centering
\caption{Fit times for analyses using \pyhf{}'s NumPy backend and SciPy optimizer orchestrated with \funcX{} on RIVER with an endpoint configuration of and \texttt{max\_blocks} = 4 and \texttt{nodes\_per\_block} = 1 over 10 trials compared to a single RIVER node. The reported wall fit time is the mean wall fit time of the trials. The uncertainty on the mean wall time corresponds to the standard deviation of the wall fit times.}
\label{table:performance}
\begin{tabular}{@{}lrrrr@{}}
\toprule
                      Analysis & Patches & Wall time (sec) & Single node (sec) \\
\midrule
 Eur. Phys. J. C 80 (2020) 691 &     125 &   $156.2\pm9.5$ &              3842 \\
             JHEP 06 (2020) 46 &      76 &    $31.2\pm2.7$ &               114 \\
Phys. Rev. D 101 (2020) 032009 &      57 &    $57.4\pm5.2$ &               612 \\
\bottomrule
\end{tabular}
\end{table}

\begin{figure}[!htpb]
    \centering
    \includegraphics[width=0.8\textwidth]{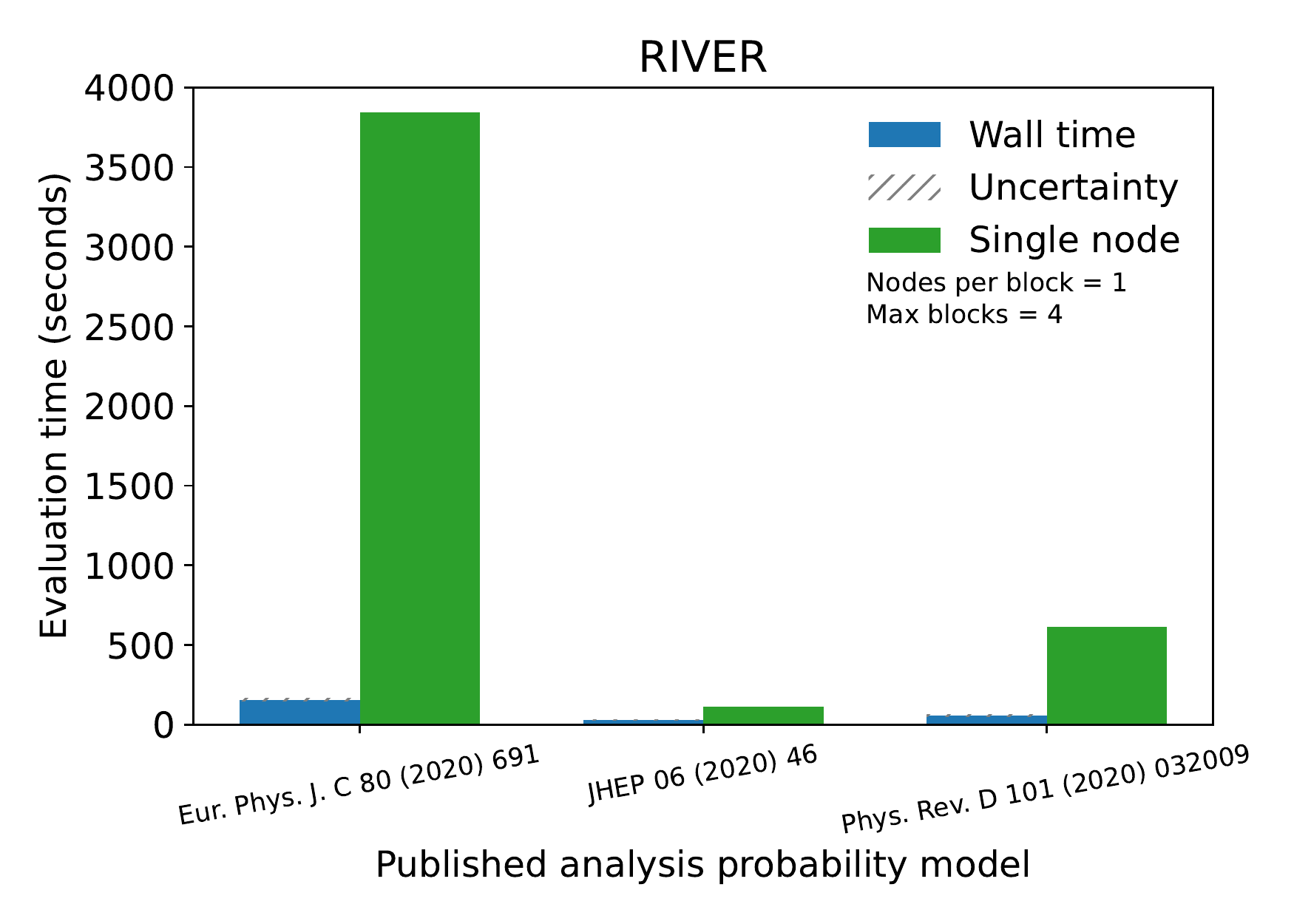}
    \caption{Visualization of comparison of the reported wall times in~\Cref{table:performance} categorized by analysis probability model for fits distributed across nodes compared to a single node.}
    \label{fig:timing_barplot_river}
\end{figure}

These results are not fundamental limits of the performance of the software and are meant as preliminary tests of scaling on heterogeneous architecture.
For comparison, on a local system with an AMD Ryzen 9 3900X processor (12 cores 3.8GHz) and 2 x 32GB DDR4-2400 Memory (64 GB) the fitting results for the 125 signal patches of Ref.~\cite{ATLAS_SUSY_1Lbb_pallet} on a single core were obtained in 1672~seconds.
Additionally, in isolated tests on RIVER the 125 signal patches of Ref.~\cite{ATLAS_SUSY_1Lbb_pallet} were able to be fit with \funcX{} orchestration in 76~seconds.
These hardware and block scaling results are parts of ongoing studies to profile the scaling performance of \funcX{} and \pyhf{} for benchmark physics analyses on additional hardware architectures at target HPC facilities.

\section{Conclusions}\label{sec:conclusions}

Through the combined use of the pure-Python libraries \funcX{} and \pyhf{}, we have demonstrated the ability to parallelize and accelerate statistical inference of physics analyses on HPC systems through a FaaS solution.
Without having to write any bespoke batch jobs, inference can be registered and executed by analysts with a client Python API that still achieves the large performance gains compared to single node execution that is a typical motivation of use of batch systems.
There is ongoing work to better monitor and extract the time costs associated with overhead and communication from the time devoted purely to statistical inference.
Characterizing these costs will allow for better understanding of the scaling behavior observed across blocks.
The results obtained on CPU further motivate the study of scaling performance with \funcX{} across GPU --- leveraging \pyhf{}'s hardware accelerated computational backends --- and the consideration of dedicated FaaS analysis facilities on HPC sites.
These additional resources have the potential to offer even further speedup through acceleration and scaling in situations where complex analyses can have individual models take over ten minutes to fit and might have multiple hundreds of model hypotheses.
The results additionally motivate investigation of the scaling performance for large scale ensemble fits in the case of statistical combinations of analyses and large dimensional scans of theory parameter space (e.g. phenomenological minimal supersymmetric standard model (pMSSM) scans)~\cite{SUSY-2014-08,Ambrogi:2017lov}.

\begin{listing}
\begin{minted}{text}
$ time python fit_analysis.py --config-file config/1Lbb.json
prepare:  waiting-for-nodes
--------------------
<pyhf.workspace.Workspace object at 0x7efbd9d95530>
Task C1N2_Wh_hbb_1000_0 complete, there are 1 results now
Task C1N2_Wh_hbb_1000_100 complete, there are 2 results now
Task C1N2_Wh_hbb_1000_150 complete, there are 3 results now
Task C1N2_Wh_hbb_1000_200 complete, there are 4 results now
Task C1N2_Wh_hbb_1000_250 complete, there are 5 results now
Task C1N2_Wh_hbb_1000_300 complete, there are 6 results now
Task C1N2_Wh_hbb_1000_350 complete, there are 7 results now
Task C1N2_Wh_hbb_1000_400 complete, there are 8 results now
Task C1N2_Wh_hbb_1000_50 complete, there are 9 results now
Task C1N2_Wh_hbb_150_0 complete, there are 10 results now
Task C1N2_Wh_hbb_165_35 complete, there are 11 results now
Task C1N2_Wh_hbb_175_0 complete, there are 12 results now
Task C1N2_Wh_hbb_175_25 complete, there are 13 results now
Task C1N2_Wh_hbb_190_60 complete, there are 14 results now
Task C1N2_Wh_hbb_200_0 complete, there are 15 results now
Task C1N2_Wh_hbb_200_25 complete, there are 16 results now
...
... skipping forward for space
...
Task C1N2_Wh_hbb_800_50 complete, there are 115 results now
Task C1N2_Wh_hbb_900_0 complete, there are 116 results now
Task C1N2_Wh_hbb_900_100 complete, there are 117 results now
Task C1N2_Wh_hbb_900_150 complete, there are 118 results now
Task C1N2_Wh_hbb_900_200 complete, there are 119 results now
inference: running
Task C1N2_Wh_hbb_900_300 complete, there are 120 results now
Task C1N2_Wh_hbb_900_350 complete, there are 121 results now
Task C1N2_Wh_hbb_900_400 complete, there are 122 results now
Task C1N2_Wh_hbb_900_50 complete, there are 123 results now
Task C1N2_Wh_hbb_535_400 complete, there are 124 results now
Task C1N2_Wh_hbb_900_250 complete, there are 125 results now
--------------------
... skipping print of results

real	2m20.750s
user	0m11.724s
sys	0m2.018s
\end{minted}
 \caption{A subset of the run output from the execution of fitting the 125 signal hypothesis patches for the published ATLAS analysis~\cite{SUSY-2019-08}.
 The wall time (\texttt{real}) shows the simultaneous fit orchestrated by \funcX{} is performed in 2 minutes and 20 seconds.}
 \label{lst:funcX_demo_output}
\end{listing}

\section{Acknowledgments}\label{sec:acknowledgements}

The authors would like to thank everyone in the Scikit-HEP developer community and the Institute for Research and Innovation in Software for High Energy Physics for their continued support and feedback.
The authors thank Sinclert P{\'e}rez for originally producing the images used to compose~\Cref{fig:infrastructure_perspective}.
Matthew Feickert and Ben Galewsky were supported by the National Science Foundation under Cooperative Agreement OAC-1836650 for this work.

\clearpage
% Don't use \bibliographystyle
% the style is already called in woc.bst, so ensure it is in the top level directory
\bibliography{bib/ref,bib/ATLAS_papers,bib/PubNotes}

\begin{thebibliography}{26}

\bibitem{ThetaANL}
\emph{{Argonne Leadership Computing Facility: Theta/ThetaGPU Machine
  Overview}},
  \url{https://www.alcf.anl.gov/support-center/theta/theta-thetagpu-overview}
  (2021), accessed: 2021-02-28

\bibitem{pyhf}
L.~Heinrich, M.~Feickert, G.~Stark, \emph{{pyhf: v0.6.0}},
  \urlstyle{tt}\url{https://doi.org/10.5281/zenodo.1169739}

\bibitem{pyhf_joss}
L.~Heinrich, M.~Feickert, G.~Stark, K.~Cranmer, Journal of Open Source Software
  \textbf{6}, 2823 (2021)

\bibitem{funcX_paper}
R.~Chard, Y.~Babuji, Z.~Li, T.~Skluzacek, A.~Woodard, B.~Blaiszik, I.~Foster,
  K.~Chard, \emph{FuncX: A Federated Function Serving Fabric for Science}, in
  \emph{Proceedings of the 29th International Symposium on High-Performance
  Parallel and Distributed Computing} (Association for Computing Machinery, New
  York, NY, USA, 2020), HPDC '20, p. 65–76, ISBN 9781450370523,
  \urlstyle{tt}\url{https://doi.org/10.1145/3369583.3392683}

\bibitem{Cranmer:1456844}
K.~Cranmer, G.~Lewis, L.~Moneta, A.~Shibata, W.~Verkerke, Tech. Rep.
  CERN-OPEN-2012-016 (2012),
  \urlstyle{tt}\url{https://cds.cern.ch/record/1456844}

\bibitem{HIGG-2013-02}
{ATLAS Collaboration}, Phys. Lett. B \textbf{726}, 88 (2013)

\bibitem{Aaij:2015sqa}
{LHCb Collaboration}, Phys. Rev. D \textbf{92}, 032002 (2015)

\bibitem{SUSY-2016-10}
{ATLAS Collaboration}, JHEP \textbf{06}, 107 (2018), \texttt{1711.01901}

\bibitem{Alguero:2020grj}
G.~Alguero, S.~Kraml, W.~Waltenberger (2020), \texttt{2009.01809}

\bibitem{Cowan:2010js}
G.~Cowan, K.~Cranmer, E.~Gross, O.~Vitells, Eur. Phys. J. C \textbf{71}, 1554
  (2011), [Erratum: Eur.Phys.J.C 73, 2501 (2013)], \texttt{1007.1727}

\bibitem{ATL-PHYS-PUB-2019-029}
{ATLAS Collaboration}, {ATL-PHYS-PUB-2019-029} (2019),
  \urlstyle{tt}\url{https://cds.cern.ch/record/2684863}

\bibitem{Maguire:2017ypu}
E.~Maguire, L.~Heinrich, G.~Watt, J. Phys. Conf. Ser. \textbf{898}, 102006
  (2017)

\bibitem{Parsl_paper}
Y.~Babuji, A.~Woodard, Z.~Li, D.S. Katz, B.~Clifford, R.~Kumar, L.~Lacinski,
  R.~Chard, J.M. Wozniak, I.~Foster et~al., \emph{Parsl: Pervasive Parallel
  Programming in Python}, in \emph{Proceedings of the 28th International
  Symposium on High-Performance Parallel and Distributed Computing}
  (Association for Computing Machinery, New York, NY, USA, 2019), HPDC '19, p.
  25–36, ISBN 9781450366700,
  \urlstyle{tt}\url{https://doi.org/10.1145/3307681.3325400}

\bibitem{IRIS-HEP:strategic-plan}
P.~Elmer, M.~Neubauer, M.D. Sokoloff (2017), \texttt{1712.06592}

\bibitem{RIVER_HPC}
\emph{{Research Infrastructure to explore Volatility, Energy-efficiency, and
  Resilience (RIVER): Usage Models and Resources}},
  \url{http://river.cs.uchicago.edu/website-builder.html} (2021), accessed:
  2021-02-28

\bibitem{SUSY-2019-08}
{ATLAS Collaboration}, Eur. Phys. J. C \textbf{80}, 691 (2020),
  \texttt{1909.09226}

\bibitem{portable_inference_workshop}
M.~Feickert, \emph{{Fitting and Statistical Inference as a Service}} (2020),
  {IRIS-HEP Blueprint Workshop on Portable Inference},
  \urlstyle{tt}\url{https://indico.cern.ch/event/972791/contributions/4121109/}

\bibitem{ATLAS_SUSY_1Lbb_pallet}
{ATLAS Collaboration}, \emph{{Search for direct production of electroweakinos
  in final states with one lepton, missing transverse momentum and a Higgs
  boson decaying into two \(b\)-jets in \(pp\) collisions at \(\sqrt{s} =
  13\,\text{TeV}\) with the ATLAS detector}} (2020),
  \urlstyle{tt}\url{https://doi.org/10.17182/hepdata.90607.v4}

\bibitem{SUSY-2018-09}
{ATLAS Collaboration}, JHEP \textbf{06}, 046 (2020), \texttt{1909.08457}

\bibitem{SUSY-2018-04}
{ATLAS Collaboration}, Phys. Rev. D \textbf{101}, 032009 (2020),
  \texttt{1911.06660}

\bibitem{ATLAS_SUSY_SS3L_pallet}
{ATLAS Collaboration}, \emph{{Search for squarks and gluinos in final states
  with same-sign leptons and jets using \(139\,\text{fb}^{-1}\) of data
  collected with the ATLAS detector}} (2020),
  \urlstyle{tt}\url{https://doi.org/10.17182/hepdata.91214.v4}

\bibitem{ATLAS_SUSY_staus_pallet}
{ATLAS Collaboration}, \emph{{Search for direct stau production in events with
  two hadronic \(\tau\)-leptons in \(\sqrt{s} = 13\,\text{TeV}\) \(pp\)
  collisions with the ATLAS detector}} (2020),
  \urlstyle{tt}\url{https://doi.org/10.17182/hepdata.92006.v2}

\bibitem{study_code}
M.~Feickert, L.~Heinrich, G.~Stark, B.~Galewsky, \emph{{Distributed Inference
  with pyhf and funcX}}, vCHEP 2021 release,
  \urlstyle{tt}\url{https://github.com/matthewfeickert/distributed-inference-with-pyhf-and-funcX}

\bibitem{study_code_zenodo_doi}
M.~Feickert, L.~Heinrich, G.~Stark, B.~Galewsky,
  \emph{matthewfeickert/distributed-inference-with-pyhf-and-funcx} (2021),
  \urlstyle{tt}\url{https://doi.org/10.5281/zenodo.4945694}

\bibitem{SUSY-2014-08}
{ATLAS Collaboration}, JHEP \textbf{10}, 134 (2015), \texttt{1508.06608}

\bibitem{Ambrogi:2017lov}
F.~Ambrogi, S.~Kraml, S.~Kulkarni, U.~Laa, A.~Lessa, W.~Waltenberger, Eur.
  Phys. J. C \textbf{78}, 215 (2018), \texttt{1707.09036}

\end{thebibliography}
\end{document}